\documentclass[aps,prl,preprint,a4paper,nofootinbib,raggedbottom]{revtex4-1}
\usepackage[autostyle]{csquotes}
\usepackage{placeins}
\usepackage{cancel}
\usepackage{hyperref}
\usepackage{amsmath}
\usepackage{amssymb}
\usepackage{amsthm}
\usepackage{graphicx}
\usepackage{dcolumn}
\usepackage{bm}
\usepackage{upgreek}
\usepackage{esint}
\usepackage{subcaption}
\usepackage{xcolor}
\usepackage[english]{babel}

\theoremstyle{definition}
\newtheorem{definition}{Definition}

\theoremstyle{remark}

\theoremstyle{plain}

\theoremstyle{plain}

\theoremstyle{plain}

\theoremstyle{plain}

\begin{document}

\title{Is There an Exact Magnetic-Moment for Charged Particle Motion in a Homogeneous, Time-Dependent Magnetic Field?}.

\author{Michael Updike}
\affiliation{Princeton Plasma Physics Laboratory, Princeton University, Princeton,
NJ 08540}
\affiliation{Department of Astrophysical Sciences, Princeton University, Princeton,
NJ 08540}
\author{J. W. Burby}
\affiliation{Department of Physics and Institute for Fusion Studies, The University of Texas at Austin, Austin, TX 78712, USA}
\date{\today}
\begin{abstract}
The nonperturbative guiding-center model provides an exact alternative to full-orbit simulations of charged particle dynamics in situations where traditional guiding-center theory may fail. We demonstrate that the charged particle motion in a homogeneous, time-varying magnetic field is a solvable example of the nonperturbative guiding-center model.
 This entails showing that the exact magnetic moment of Qin and Davidson can be constructed to be asymptotic to the adiabatic invariant series of Kruskal. In contrast to the perturbative invariant, the exact invariant contains information about parametric resonances. These resonances destroy the conservation of the usual magnetic moment over very long times. This refutes some previous claims about the all-time invariance of the magnetic moment.

\end{abstract}
\maketitle 
\newpage
\section{Introduction}
Traditional models of particle motion in strong magnetic fields rely on Kruskal's asymptotic theory of nearly-periodic systems \cite{Kruskal_1962} to formally separate the fast gyration of charged particles from slow drift motions. Examples include the guiding-center \cite{kruskal58, Littlejohn_1984,10.1063/1.863594,Carey} and gyrocenter \cite{gyrocenter1} models. For Hamiltonian theories, such as the aforementioned, Kruskal's theory implies the existence of a formally conserved adiabatic invariant series $\mu_{\epsilon} = \mu_{0} + \epsilon \mu_{1} + \epsilon^2 \mu_{2} + \hdots$. Consequently, traditional models of single-particle motion are accurate when $\mu_{\epsilon}$ is well conserved, but suffer when the particle Larmor radius (cyclotron period) is not particularly small compared to the length (time) scale of magnetic field variation. These models may therefore lead to imprecise estimates of particle confinement, particularly in device regions where the magnetic field varies sharply. This is a primary concern for the confinement of energetic $\alpha$-particles. The $\alpha$-particle gyro-radii are only marginally smaller than the length-scale of magnetic field variation in typical stellarators. As recounted in \cite{j_w_burby_nonperturbative_2025}, significant deviations in the predicted $\alpha$-particle loss fraction between guiding-center and full-orbit simulations may therefore occur in certain magnetic geometries \cite{numerics1,numerics2,numerics3}.

These concerns motivated Burby et al. \cite{j_w_burby_nonperturbative_2025} to develop a nonperturbative guiding-center model. The nonperturbative guiding center model postulates the existence of an exact invariant $\mathcal{J}_{\epsilon}$, dubbed the nonperturbative adiabatic invariant, which both generates a $U(1)$ symmetry, and agrees with $\mu_{\epsilon}$ to at all orders in $\epsilon$. Assuming such a $\mathcal{J}_{\epsilon}$ can be found, the nonperturbative guiding center theory gives exact equations for charged particle motion free of the fast cyclotron timescale. Further, the nonperturbative guiding-center model recovers the usual guiding-center theory to all orders in $\epsilon$. No examples of nonperturbative adiabatic invariants were given in \cite{j_w_burby_nonperturbative_2025}. Rather, Burby et al. demonstrated that approximate nonperturbative adiabatic invariants can be constructed from data. The particle dynamics recovered from these learned invariants outperformed the traditional guiding center model. This result motivates the search for true nonperturbative adiabatic invariants. 

Recently, Hollas et al. \cite{Hollas} were able to find exact formulas for nonperturbative adiabatic invariants in some symmetric, time-independent magnetic fields. In this work, we continue the program of Hollas et al. by showing that there is a nonperturbative adiabatic invariant for particle motion in a homogeneous, time-dependent magnetic field. This system is already known to be integrable. 
In \cite{qin_exact_2006}, Qin and Davidson demonstrated that charged particle motion in a magnetic field of the form $\mathbf{B} = B(t)\hat{z}$ possesses a nontrivial, time-dependent invariant $\mathcal{M}_{\epsilon}$. The construction of Qin and Davidson follows closely the construction of H.R. Lewis \cite{Lewis:1968yx}, who showed that the time-dependent simple harmonic oscillator, henceforth referred to as Hill's equation\footnote{Usually the name \enquote{Hill's equation} is reserved for periodic $\Omega(t)$.},
\begin{equation}
    \epsilon^2 y'' + \Omega^2(t) y = 0,
\end{equation}
possesses a time-dependent invariant of motion. Parameterizing either invariant is a solution branch $w(t;\epsilon)$ of the singularly perturbed Ermakov-Pinney equation 
\begin{equation}\label{EPeqn1}
    \epsilon^2 w''(t;\epsilon) + \Omega^2(t) w(t;\epsilon) = \frac{1}{w(t;\epsilon)^3}.
\end{equation}
Taking $\epsilon$ formally to zero, Qin and Davidson showed the solution $w(t;0) = \Omega^{-1/2}(t)$ reduces $\mathcal{M}_{0}=\frac{v_{\perp}^2}{2B(t)}$ to the usual magnetic moment. For $\mathcal{M}_{\epsilon}$ to comprise a true nonperturbative adiabatic invariant, it must show that $\mathcal{M}_{\epsilon}$ agrees with $\mu_{\epsilon}$ to all orders in $\epsilon$, and that the Hamiltonian flow of $\mathcal{M}_{\epsilon}$ generates a $U(1)$ action. This has not been previously shown. 

A complication to showing all orders agreement between $\mathcal{M}_{\epsilon}$ and $\mu_{\epsilon}$ is that there is generally no solution to Eqn. \eqref{EPeqn1} taking the form of a convergent power series $w(t;\epsilon) = w_{0}(t) + \epsilon^2 w_{2}(t) + \epsilon^4 w_{4}(t) + \hdots$. It is not even clear if $w(t;\epsilon)$ can be constructed to admit an asymptotic power series. For periodic magnetic fields, for example, the formal solution $w_{0}(t) + \epsilon^2 w_{2}(t) + \hdots$ is periodic to all orders. However, we will show that there are generally arbitrarily small $\epsilon$ values where no periodic solution to Eqn. \eqref{EPeqn1} exists. At these $\epsilon$ values, particle trajectories grow exponentially unbounded. This behavior was not explored in Qin and Davidson, since no careful exploration of the singularly perturbed Ermakov-Pinney equation was conducted. This complication led Qin and Davidson to erroneously conclude that the existence of an exact invariant $\mathcal{M}_{\epsilon}$ implies the usual magnetic moment is conserved up to $O(\epsilon)$ error for $\textit{all}$ times. 

Despite this complication, we will show that the invariant $\mathcal{M}_{\epsilon}$ can be constructed to be a nonperturbative adiabatic invariant. This result follows after proving two facts. The first is that $\mathcal{M}_{\epsilon}$ generates a $U(1)$ action. The second is that there exists a branch of solutions $w(t;\epsilon)$ to the singularly-perturbed Ermakov-Pinney equation 
such that both $w(t;\epsilon)$ and $w'(t;\epsilon)$ admit asymptotic power series valid over, at least, $O(1)$ times. After proving $\mathcal{M}_{\epsilon}$ is a nonperturbative adiabatic invariant, we explore how the nonperturbative dynamics predicted by the invariance of $\mathcal{M}_{\epsilon}$ differs from the approximate dynamics predicted by the invariance of $\mu_{\epsilon}$. As we will explore, the phenomena of parametric resonance can exist for any $\epsilon > 0$.  This presents an unexplored, strictly nonperturbative phenomenon which can be inferred from the invariance of $\mathcal{M}_{\epsilon}$, but not from the approximate conservation of $\mu_{\epsilon}$ to any order in $\epsilon$.

One of the distinguishing features of this work is that the guiding-center scaling is not used to place the equations of motion in a form amenable to Kruskal's theory of nearly-periodic systems. Rather, gryocenter scaling is used. This difference is interesting but ultimately of little importance since the nonperturbative guiding-center theory of Burby et al. can be adopted to any nearly-periodic Hamiltonian system. The more important difference between this work and that of Hollas et al. \cite{Hollas} is that the Arnold-Liouville theorem cannot be leveraged in proving $\mathcal{M}_{\epsilon}$ is a nonperturbative adiabatic invariant. The analysis in this work is therefore qualitatively different from previous works.

In section \hyperref[sec1]{I}, we briefly review Kruskal's theory of adiabatic invariants in nearly-periodic Hamiltonian systems and define precisely the notion of a nonperturbative adiabatic invariant. We then show that the magnetic system of Qin and Davidson can be given the form of a nearly-periodic Hamiltonian system. We conclude by showing that $\mathcal{M}_{\epsilon}$ generates a $U(1)$ action for any choice of solution to Eqn. \eqref{EPeqn1}. In section \hyperref[sec2]{II}, we show that the Ermakov-Pinney equation is controlled by Hill's equation. We then use the Bremmer series to construct a smooth choice of $\mathcal{M}_{\epsilon}$ admitting an asymptotic series. In section \hyperref[sec3]{III}, we focus on periodic magnetic fields and use Floquet theory to explore the nonperturbative dynamics of the charged particle motion. In particular, we show that long-term particle confinement cannot be predicted from perturbative theory alone. In section \hyperref[sec4]{IV}, we give an example and explore the validity of perturbative theory numerically. Finally, we conclude with some remarks and avenues of further work in section \hyperref[discussion]{V}.

\section{Section I} \label{sec1}
Consider the non-relativistic motion of a particle of mass $m$ and charge $q = |q| \sigma$ in a uniform, time-dependent magnetic field of the form $\mathbf{B}(t) = B_{0} B(t/T) \hat{z}$. $B(t/T)$ is a dimensionless function assumed to be strictly positive, $B_{0}$ is the typical magnetic field strength, and $T$ is the characteristic timescale of magnetic field variation. We define the dimensionless ordering parameter $\epsilon^{-1} = \omega_{c}T$ where $\omega_{c} = \frac{|q|B_{0}}{m}$ is the typical cyclotron frequency. We adopt units of time such that $T = 1$. The single-particle equations of motion on the fast timescale $\tau = \epsilon^{-1}t$ are 
\begin{equation}\label{EOM}
\frac{d^2 \mathbf{x}}{d\tau^2} =  \sigma \left[B(\epsilon \tau) \frac{d \mathbf{x}}{d \tau} \times \hat{z} + \epsilon \mathbf{E}(\mathbf{x},\epsilon \tau) \right],
\end{equation}
where the induced electric field is given in cylindrical coordinates by $\mathbf{E}(\mathbf{x},t) = -\frac{1}{2} r B'(t) \hat{\theta}$\footnote{We implicitly assume $B(t)$ is generated by a long, rotationally symmetric solenoid.}. 
Since $z'' = 0$, we ignore the trivial motion in the $z$ direction. Given the rotational symmetry of Eqn. \eqref{EOM}, it is clear that polar coordinates may be used to simplify the dynamics. Defining the canonical momenta $p_{r} = \frac{d r}{d \tau}$ and $p_{\theta} = r^2 \frac{d \theta}{d \tau} + \frac{r^2 \sigma B(\epsilon \tau)}{2} $, the Hamiltonian of Eqn. \eqref{EOM} is
\begin{equation}
H = \frac{1}{2}\left(p_{r}^2 + \left( \frac{p_{\theta}}{r} - r \sigma \Omega(\epsilon \tau) \right)^2 \right),
\end{equation} 
where $\Omega = \frac{B}{2}$.

To study invariants of time-dependent systems, it is necessary to extend the phase space to include $(\tau, p_{\tau})$ as a canonically conjugate position-momentum pair obeying $\dot{\tau} = 1$. Using non-canonical coordinates, the Hamiltonian in the extended phase space can be taken to be 
\begin{equation}\label{Hamiltonian}
    \tilde{H}_{\epsilon} = \frac{\epsilon}{2}\left(p_{r}^2 + \left( \frac{p_{\theta}}{r} - r \sigma \Omega(t) \right)^2 \right) + \epsilon p_{\tau},
\end{equation}
where the symplectic form is $-d\vartheta_\epsilon$ and the Liouville one-form is
\begin{equation}\label{sympform}
    \vartheta_{\epsilon} = \epsilon p_{r}dr  + \epsilon p_{\theta}d\theta +p_{\tau} dt .
\end{equation}
The equations of motion are 
\begin{flalign}\label{HamSystem}
&\dot{p_{\theta}} = 0, \\
&\dot{\theta} = \left(\frac{p_{\theta}}{r^2} - \sigma \Omega(t) \right), \nonumber \\
&\dot{p_{\tau}} = \epsilon \sigma \Omega'(t) \left(p_{\theta} - r^2 \sigma \Omega(t) \right), \nonumber \\
&\dot{t} = \epsilon, \nonumber \\
&\dot{p_{r}} = \frac{p_{\theta}^2}{r^3} - \Omega^2(t) r, \nonumber \\
&\dot{r} = p_{r}. \nonumber 
\end{flalign}  
In the variables $(r, \theta, t, p_{r}, p_{\theta}, p_{\tau})$, Eqn. \eqref{HamSystem} takes the desired form of a nearly-periodic Hamiltonian system. As shown in the seminal work of Kruskal \cite{Kruskal_1962}, nearly-periodic Hamiltonian systems encompass a broad class of systems for which a perturbative adiabatic invariant $\mu_{\epsilon} = \mu_{0} + \epsilon \mu_{1} + \hdots$ can be constructed, and therefore provide a general context for defining the notion of a nonperturbative adiabatic invariant. For the sake of completeness, we recall the generalized definition of a nearly periodic system as defined by Burby and others in \cite{burby_general_2020, burby_nearly_periodic_maps, normal_stab}:
\begin{definition}
    \textit{Nearly periodic Hamiltonian system}. A nearly periodic system on an (possibly infinite-dimensional) manifold $M$ is an ODE of the form $\dot{z} = X_{\epsilon}(z)$ where 
    \begin{itemize}
        \item $X_{\epsilon}$ is a smooth vector field depending smoothly on $\epsilon$ for $0 \leq \epsilon \ll 1$. 
        \item There exists a smooth, positive function $\omega_{0}$ and a vector field $R_{0}$ such that $X = \omega_{0} R_{0}$. 
        \item $R_{0}$ generates a $2\pi$-periodic $U(1)$ action.
        \item $\mathcal{L}_{R_{0}} \omega_{0} = 0$.
    \end{itemize}
    If, in addition, there exists a smoothly parameterized family of  functions $H_{\epsilon}$ and one-forms $\vartheta_{\epsilon}$ such that
    \begin{itemize}
        \item $i_{X_{\epsilon}} d \vartheta_{\epsilon} = -dH_{\epsilon}$, 
    \end{itemize}
    then we say that $(X_{\epsilon}, \vartheta_{\epsilon}, H_{\epsilon})$ is a nearly periodic Hamiltonian system. 
\end{definition}
Critical to the construction of an adiabatic invariant $\mu_{\epsilon}$ in nearly-periodic Hamiltonian systems is the roto-rate vector $R_{\epsilon} = R_{0} + \epsilon R_{1} + \hdots$ which is the unique formal vector field satisfying, in the sense of formal power series, 
\begin{itemize}\label{Roto-rate condition}
\begin{item}
    $\displaystyle\lim_{\epsilon \to 0} R_{\epsilon} = R_{0}$.
\end{item}
\begin{item}
    $\exp(2 \pi R_{\epsilon}) = \text{Id}$. 
\end{item}
\begin{item}
    $[X_{\epsilon}, R_{\epsilon}] =0 $.
\end{item}
\end{itemize}
It is not hard to show that the conditions imposed on the roto-rate vector ensure $R_{\epsilon}$ is a formal Noether symmetry. The asymptotic invariant $\mu_{\epsilon} = \mu_{0} + \epsilon \mu_{1} + \hdots$ is then the associated formally conserved quantity obeying $i_{R_{\epsilon}} d \vartheta_{\epsilon} = -d \mu_{\epsilon}$.  As argued in \cite{relectionless}, even under ideal conditions, the series for $\mu_{\epsilon}$ may not converge for any $\epsilon > 0$. In general, there is no coordinate change separating, say, the fast motion of gyrating particles from the slow drift motion. The results of Neishtadt and others \cite{the_sep_of_motions,gevsep,averaging_under_QP_forcing} show that such a split may only be possible up to an exponentially small coupling between the fast and slow variables.

While it is too much to ask that the formal series for $R_{\epsilon}$ converges, it may nevertheless be true that a $U(1)$ symmetry persists for $\epsilon > 0$. In such a case, we say the symmetry generator $\mathcal{R}_{\epsilon}$ is a nonperturbative roto-rate vector if $\mathcal{R}_{\epsilon}$ depends continuously on  $0\leq \epsilon \ll 1$ and admits an asymptotic power series as $\epsilon \to 0^{+}$. Uniqueness of the roto-rate then implies $\mathcal{R}_{\epsilon} \sim R_{\epsilon}$. When a nonperturbative roto-rate vector exists in a nearly periodic Hamiltonian system, the Hamiltonian $\mathcal{J}_{\epsilon}$ generating $\mathcal{R}_{\epsilon}$ is a nonperturbative adiabatic invariant. Indeed, since $\mathcal{R}_{\epsilon}$ is asymptotic to $R_{\epsilon}$, it follows that $\mathcal{J}_{\epsilon}$ is asymptotic to Kruskal's adiabatic invariant $\mu_{\epsilon}$. This leads to the following definition:

\begin{definition}\label{nonperturbative}
    \textit{Nonperturbative Adiabatic Invariant}. Given a nearly-periodic Hamiltonian system $(X_{\epsilon}, \vartheta_{\epsilon}, H_{\epsilon})$, we say that an $\epsilon$-parameterized family of functions $\mathcal{J}_{\epsilon}$ is a nonperturbative adiabatic invariant if 
    \begin{itemize}
        \item $\mathcal{L}_{X_{\epsilon}} \mathcal{J}_{\epsilon} = 0$.
        \item $\mathcal{J}_{\epsilon}$ is continuous in $\epsilon$ for $0 \leq \epsilon \ll 1$.
        \item $\mathcal{J}_{\epsilon}$ admits an asymptotic power series of the form $\mathcal{J}_{\epsilon} \sim J_{0} + \epsilon J_{1} + \hdots$  as $\epsilon \to 0^{+}$.
        \item There exists a family of vector fields $\mathcal{R}_{\epsilon}$ such that $i_{\mathcal{R}_{\epsilon}} d \vartheta_{\epsilon} = -d \mathcal{J}_{\epsilon}$. 
        \item $\displaystyle \lim_{\epsilon \to 0} \mathcal{R}_{\epsilon} = R_{0}$.
        \item $\exp(2 \pi \mathcal{R}_{\epsilon}) = \text{Id}$. 
    \end{itemize}
\end{definition}

In the present case,  $\tilde{H}_{\epsilon}$ and $\vartheta_{\epsilon}$ are given by Eqns. \eqref{Hamiltonian} and $\eqref{sympform}$ respectively, and $X_{\epsilon}$ is the Hamiltonian vector field \eqref{HamSystem}. Taking the formal limit $\epsilon \to 0$, the limiting dynamics $X_{0}$ consists of periodic motion at an angular frequency of $2\Omega(t)$. We therefore  have a nearly-periodic Hamiltonian system with $\omega_{0}(t) = 2 \Omega(t)$ and $R_{0} = \frac{X_{0}}{2 \Omega(t)}$. 

In Qin and Davidson \cite{qin_exact_2006}, it was shown that 
\begin{equation}
2\epsilon^{-1} \mathcal{M}_{\epsilon} = \frac{ p_{\theta}^2}{2} \left(\frac{w(t; \epsilon)^2}{r^2} \right) + \frac{1}{2} \left(\frac{r^2}{w(t;\epsilon)^2} \right) + \frac{1}{2}(p_{r}w(t;\epsilon) - \epsilon w'(t;\epsilon)r)^2 - p_{\theta}\sigma,
\end{equation}
is an exact invariant of the dynamics when $w(t;\epsilon)$ solves the singularly perturbed Ermakov-Pinney equation
\begin{equation}\label{pinneyeqn}
    \epsilon^2 w''(t;\epsilon) = \frac{1}{w^3(t;\epsilon)} - \Omega^2(t) w(t;\epsilon).
\end{equation}
Taking the formal limit $\epsilon \to 0$, one computes that $w(t;0) = \Omega(t)^{-1/2}$. Thus, $\epsilon^{-1}\mathcal{M}_{\epsilon}$ formally limits to magnetic moment $\epsilon^{-1} \mathcal{M}_{0} = \frac{1}{4 \Omega}\left( r^2 \dot{\theta}^2 + p_{r}^2\right) = \mu_{1}$, and the Hamiltonian vector field of $\mathcal{M}_{\epsilon}$ limits to $R_{0}$. To prove $\mathcal{M}_{\epsilon}$ is a nonperturbative adiabatic invariant, it must be shown that $\mathcal{M}_{\epsilon}$ generates a $2 \pi$-periodic $U(1)$ action, and that there exists a branch of solutions $w(t; \epsilon)$ to the Ermakov-Pinney equation such that both $w(t;\epsilon)$ and $w'(t;\epsilon)$ admit asymptotic power series in $\epsilon$. Once both of these statements have been shown, one verifies all the conditions of Def. \ref{nonperturbative} are met. 

We first assume $w(t;\epsilon)$ is given and show that $\mathcal{M}_{\epsilon}$ generates a $U(1)$ action. In the next section, we show a method by which $w(t;\epsilon)$ may be smoothly constructed. 

The Hamiltonian vector field of $2\mathcal{M}_{\epsilon}$ generates the flow
\begin{align}
    &\frac{d t}{d \xi}   = 0, \\
    &\frac{d p_{\tau}}{d \xi} = \epsilon \left[-ww'\left(\frac{p_{\theta}^2}{r^2} + p_{r}^2 \right) + r^2 w'\left(\frac{1}{w^3} - \epsilon^2 w''\right) + rp_{r}\epsilon \left(w''w + w'^2\right)\right], \nonumber \\
    &\frac{d r}{d \xi} =  p_{r}w^2 - \epsilon w w' r, \nonumber \\
   & \frac{d p_{r}}{d \xi}  = \frac{p_{\theta}^2 w^2}{r^3} - \frac{r}{w^2} + \epsilon w' \left(p_{r} w - \epsilon w' r \right),  \nonumber \\
    &\frac{d \theta}{d\xi} = \frac{p_{\theta}w^2}{r^2} - \sigma, \nonumber \\
    &\frac{d p_{\theta}}{d \xi} = 0. \nonumber 
\end{align}
The equation for the canonical pair $(r,p_{r})$ closes to give 
\begin{equation}
    r'' + r = \frac{p_{\theta}^2 w^4}{r^3},
\end{equation}
which has the general solution $r^2(\xi; E, \phi) = p_{\theta}w^2 [E + \sqrt{E^2 - 1}\sin(2\xi + \phi)]$. Given that $p_{r}(\xi) = \frac{1}{w^2}\left(\frac{d r}{d \xi}(\xi) + \epsilon ww'r(\xi)\right)$, $p_{r}(\xi)$ is a $\pi$-periodic function for any constants $E \geq 1, \phi$. One computes that 
\begin{align}
&\int_{0}^{\pi} \frac{d\xi}{r^2(\xi)} = \frac{\pi}{p_{\theta} w^2}, \\
    &\int_{0}^{\pi}  r^2(\xi) d\xi = (\pi p_{\theta} E)\cdot w^2, \nonumber \\
    &\int_{0}^{\pi} \left[ \frac{p_{\theta}^2}{r^2(\xi)} + (p_{r}(\xi))^2\right]d \xi = (\pi p_{\theta} E) \cdot \left( \frac{1}{w^2}+\epsilon^2 w'^2\right), \nonumber \\
    &\int_{0}^{\pi} r(\xi) p_{r}(\xi) d\xi = (\pi p_{\theta}E)\cdot ww' \epsilon. \nonumber
\end{align}
It follows that $\epsilon^{-1}\mathcal{M}_{\epsilon}$ generates a $2\pi$ periodic $U(1)$ action since
\begin{align}
    \Delta \theta & \equiv \int_{0}^{\pi} \frac{d\theta}{d \xi}(\xi) d \xi = \pi - \sigma \pi \equiv 0 \, (\text{mod } 2\pi), \\
    \Delta p_{\tau} & \equiv \int_{0}^{\pi} \frac{d p_{\tau}}{d \xi}(\xi) d \xi = 0. \nonumber 
\end{align}
We note that when a Hamiltonian system is only known to be (partially) integrable, one might appeal to generalizations of the Arnold-Liouville theorem \cite{poincare-lyapunov, non-compact-Liouville-Arnold} to ensure the existence of action-like variables.
\section{Section II}\label{sec2}
We now construct a smooth function $w(t;\epsilon)$ satisfying the singularly perturbed Ermakov-Pinney equation
\begin{equation}
    \epsilon^2 w''(t;\epsilon) + \Omega^2(t) w(t;\epsilon) = \frac{1}{w^3(t;\epsilon)},
\end{equation}
admitting an asymptotic power series in $\epsilon$ limiting to $\Omega(t)^{-1/2}$. We show that $w'(t;\epsilon)$ also admits an asymptotic power series obtained by term-by-term differentiation of the series for $w(t;\epsilon)$. Provided this choice of $w(t;\epsilon)$ is used in the construction of the exact invariant, $\mathcal{M}_{\epsilon}$ admits an asymptotic series and is hence a nonperturbative adiabatic invariant. 

It is known by the general results of Sibuya and others \cite{sibuyaI,sibuyaII,gevry_solns} that solutions to singularly perturbed analytic ODEs asymptotic to their formal series solution exist for short times. To obtain all-time results for merely smooth $\Omega(t)$, use the well-known fact that the Ermakov-Pinney equation 
\begin{equation}\label{EPeqn}
\epsilon^2 w''(t;\epsilon) + \Omega^2(t) w(t;\epsilon) = \frac{1}{w(t;\epsilon)^3}, 
\end{equation}
is controlled by the singularly perturbed Hill's equation \cite{Ermakov-Pinney-eqn-soln}
\begin{equation}\label{hillseqn}
    \epsilon^2 \psi''(t;\epsilon) + \Omega^2(t) \psi(t;\epsilon) = 0.
\end{equation}
Indeed, suppose one has found two linearly independent $u(t),v(t)$ solutions to Hill's equation. Then there exists smooth, real functions $R(t) > 0$ and  $\phi(t)>0$ such that 
\begin{equation}
\psi \equiv u(t) + i v(t) = R(t) \exp\left(\mp\frac{i}{\epsilon} \int^{t}_{0} \phi(t')dt'\right).
\end{equation}
$R$ and $\phi$ define a solution to Hill's equation if and only if
\begin{align}
    \epsilon^2R'' + \Omega^2(t)R &= \phi^{2}R, \\
    2\phi R' + \phi' R &= 0. \nonumber
\end{align}
The latter equation implies $R = C \phi^{-1/2}$. Provided the initial conditions are such that $C = 1$, which can always be ensured by rescaling $u$ and $v$, we have that 
\begin{equation}
\epsilon^2 R'' + \Omega^2(t) R = \frac{1}{R^3}.
\end{equation}
With this basic fact in mind, it is clear that the formal expansion of $w(t;\epsilon)$ into powers of $\epsilon$ is obtained from the usual WKB series of Eqn. \eqref{hillseqn}. Indeed, the first-order WKB approximation yields that $w(t;\epsilon \ll 1) \approx \Omega^{-1/2}$. Results about the accuracy of WKB \cite{Olver_1961,OLVER1974190,TAYLOR198279,asmtotitcs_and_borel_summability} therefore justifies that the formal solutions $w^{(n)}(t) = w_{0}(t) + \epsilon^{2}w_{2}(t) + \hdots + \epsilon^{n} w_{n}(t)$ to Eqn. \eqref{EPeqn} approximate some true solution for any fixed $n$.

To construct an exact solution $\psi(t;\epsilon)$ to Hill's equation asymptotic to the WKB series at all orders, we use the Bremmer series. Constructed independently in \cite{https://doi.org/10.1002/cpa.3160040111,bremmerseries2,Bremmer3} and studied further in \cite{KAY196140,ATKINSON1960255,invar_embedding}, the Bremmer series organizes the solution of Eqn. \eqref{hillseqn} into a convergent sum of waves undergoing a particular number of reflections. Importantly, the Bremmer series agrees with WKB at leading order. In Winitzki \cite{PhysRevD.72.104011}, it was shown that the Bremmer series solutions admit an asymptotic $\epsilon$-series provided there exists a point in time such that all derivatives of $\Omega(t)$ vanish. By temporarily introducing a cutoff to $\Omega(t)$ in the far past, we show that this assumption can be dropped. We repeat parts of Winitzski for the sake of a self-contained construction, but refer the reader to Winitzski \cite{PhysRevD.72.104011} for an interesting discussion on the precision of optimally truncated WKB.

We first write $\Omega(t) = \Omega_{0} + \phi(t)$ where $\phi(t)$ is smooth and the strict inequality $|\phi(t)| < \Omega_{0}$ holds uniformly. Let $0 \leq H(t) \leq 1$ be a smooth bump function supported on a domain of the form $[-K,\infty)$ such that $H(t\geq 0) \equiv 1$. We define the deformed magnetic field $\tilde{\Omega}(t) = \Omega_{0} + H(t) \phi(t)$. Clearly, for $t \geq 0$, we have not changed $\Omega(t)$ so any  linearly independent solutions to the deformed equation 
\begin{equation}\label{deformed}
    \epsilon^2\tilde{\psi}''(t;\epsilon) + \tilde{\Omega}^2(t) \tilde{\psi}(t;\epsilon) = 0,
\end{equation}
will yield a solution to the undeformed Hill's equation, hence the undeformed Ermakov-Pinney equation, for $t \geq 0$. By deforming the magnetic field, we are able to assume $\tilde{\psi}$ is a positive-frequency WKB solution in the far past, i.e. 
\begin{equation}
\tilde{\psi}(t \to -\infty;\epsilon) = \tilde{\Omega}^{-1/2}(t)\exp\left(-\frac{i}{\epsilon} \int_{0}^{t} \tilde{\Omega}(t')dt'\right).
\end{equation}
Making this notion more precise, we define the WKB solutions 
\begin{equation}
X_{\pm}(t;\epsilon) = \tilde{\Omega}^{-1/2}(t)\exp\left(\mp \frac{i}{\epsilon} \int^{t}_{0} \tilde{\Omega}(t')dt'\right).
\end{equation}
Then there exists complex-valued functions $p(t;\epsilon)$ and $q(t;\epsilon)$ such that 
\begin{equation}
\tilde{\psi}(t;\epsilon) = p(t;\epsilon)X_{+}(t;\epsilon) + q(t;\epsilon)X_{-}(t;\epsilon).
\end{equation}
The condition that $\tilde{\psi}$ is a WKB solution in the far past implies initial conditions $p(-\infty;\epsilon) = 1$ and $q(-\infty;\epsilon) = 0$. Since $p(t;\epsilon)$ and $q(t;\epsilon)$ define four degrees of freedom, we may remove two of these degrees of freedom by demanding that
\begin{equation}
    \frac{d \tilde{\psi}}{dt}(t;\epsilon) = -\frac{i \Omega(t)}{\epsilon}[p(t;\epsilon) X_{+}(t;\epsilon)  - q(t;\epsilon)X_{-}(t;\epsilon)].
\end{equation}
$\tilde{\psi}$ satisfies the deformed Hill's equation and has the desired form in the far past iff
\begin{align}\label{piccard}
    p(t;\epsilon) &= 1 + \int_{-\infty}^{t} \frac{q(t';\epsilon)}{2\tilde{\Omega}(t')} \frac{d\tilde{\Omega}}{dt}(t')\exp\left(\frac{2i}{\epsilon} \int_{0}^{t'} \tilde{\Omega}(t'')dt''\right) dt', \\
    q(t;\epsilon) &= \int_{-\infty}^{t} \frac{p(t';\epsilon)}{2 \tilde{\Omega}(t')} \frac{d \tilde{\Omega}}{dt}(t') \exp\left(-\frac{2i}{\epsilon} \int_{0}^{t'} \tilde{\Omega}(t'')dt''\right) dt'  \nonumber.
\end{align}
Eqn. \eqref{piccard} yields a convergent Picard iteration scheme when the initial guess $(p_{0},q_{0}) = (1,0)$ is assumed. Equivalent to the Piccard iteration, we may write $p(t;\epsilon) = \sum_{n = 0}^{\infty} p_{n}(t;\epsilon)$ and $q(t;\epsilon) = \sum_{n = 1}^{\infty} q_{n}(t;\epsilon)$ with $p_{0} = 1$ and with
\begin{align}\label{picit}
q_{n+1}(t;\epsilon) &=  \int_{-\infty}^{t} \frac{p_{n}(t';\epsilon)}{2\tilde{\Omega}(t')} \frac{d\tilde{\Omega}}{dt}(t')\exp\left(-\frac{2i}{\epsilon} \int_{0}^{t'} \tilde{\Omega}(t'')dt''\right) dt', \\
p_{n+1}(t;\epsilon) &=  \int_{-\infty}^{t} \frac{q_{n+1}(t';\epsilon)}{2\tilde{\Omega}(t')} \frac{d\tilde{\Omega}}{dt}(t')\exp\left(\frac{2i}{\epsilon} \int_{0}^{t'} \tilde{\Omega}(t'')dt''\right) dt'. \nonumber
\end{align}
The $\epsilon$-independent bound
\begin{equation}
\sum_{n =0 }^{\infty} |p_{n}(t;\epsilon)| + \sum_{n = 0}^{\infty} |q_{n}(t;\epsilon)| \leq \exp\left(\int_{-\infty}^{t} \left|\frac{1}{2 \tilde{\Omega}(t')} \frac{d \tilde{\Omega}}{dt}(t') \right| dt' \right) < \infty, 
\end{equation}
proves that the series for $p$ and $q$ converges uniformly on time-intervals of the form $(-\infty,T]$. To show that that $\tilde{\psi}$ admits an asymptotic series of the form $\tilde{\psi} \sim (1 + \epsilon \tilde{\psi}_{1}(t) + \hdots) X_{+}$, we define the variable $\zeta(t) = \int_{0}^{t} \tilde{\Omega}(t') dt'$ and the function $\alpha(\zeta) = \frac{1}{2 \tilde{\Omega}^2} \frac{d \tilde{\Omega}}{dt}(\zeta)$. Then 
\begin{align}
q_{n+1}(\zeta;\epsilon) &=  \int_{-\infty}^{\zeta} p_{n}(\zeta';\epsilon) \alpha(\zeta') \exp\left(-\frac{2i}{\epsilon} \zeta' \right) d\zeta', \\
p_{n+1}(\zeta;\epsilon) &=  \int_{-\infty}^{\zeta} q_{n+1}(\zeta';\epsilon) \alpha(\zeta') \exp\left(\frac{2i}{\epsilon} \zeta' \right) d\zeta'. \nonumber
\end{align}
Repeatedly integrating by parts, the standard result that $\displaystyle\lim_{\epsilon \to 0}\int_{D} \, f(\zeta') \exp\left(\pm \frac{i \zeta'}{\epsilon}\right)d \zeta'  = 0$ for any $f \in L^{1}(D\subset \mathbb{R})$ shows that
\begin{equation}
q_{1}(\zeta;\epsilon) = \int_{-\infty}^{\zeta} \alpha(\zeta') \exp\left(- \frac{2 i}{\epsilon} \zeta'\right)d \zeta' \sim -\frac{\epsilon}{2i} \exp\left(-\frac{2 i}{\epsilon} \zeta \right)\sum_{n = 0}^{\infty} \frac{\epsilon^{n}}{(2i)^{n}} \frac{d^{n}\alpha}{d \zeta^{n}} (\zeta).
\end{equation}
Since term-by-term integration of an asymptotic series yields an asymptotic series, we have that 
\begin{equation}
p_{1}(\zeta;\epsilon) = \int_{-\infty}^{\zeta} q_{1}(\zeta';\epsilon) \alpha(\zeta') \exp\left(\frac{2 i}{\epsilon} \zeta'\right) d \zeta' \sim -\frac{\epsilon}{2i}\sum_{n = 0}^{\infty} \frac{\epsilon^{n}}{(2i)^{n}} \int_{-\infty}^{\zeta} \alpha(\zeta') \left[ \frac{d^{n} \alpha}{d \zeta^{n}} (\zeta')\right]d \zeta'. 
\end{equation}
Continuing in this fashion, its not hard to see that $q_{n}$ admits an asymptotic series of the form $q_{n} \sim \epsilon^{n} \exp\left(-\frac{2i}{\epsilon}\zeta\right)\left( q_{n}^{0} + \epsilon q_{n}^{1} + \hdots \right)$ and $p_{n}$ admits an asymptotic powers series of the form $p_{n} \sim \epsilon^n (p_{n}^{0} + \epsilon p_{n}^{1} + \hdots)$. For any $N$, the finite sums $\sum_{n = 0}^{N} p_{n}$ and $\exp(\frac{2 i}{\epsilon} \zeta) \sum_{n = 1}^{N} q_{n}$ admit asymptotic power series. To see that the limiting functions $p = \sum_{n = 0}^{N} p_{n} + \sum_{n > N} p_{n}$ and $\exp(\frac{2 i}{\epsilon} \zeta) q = \exp(\frac{2 i}{\epsilon} \zeta) \sum_{n = 1}^{N} q_{n} + \exp(\frac{2 i}{\epsilon} \zeta) \sum_{n> N} q_{n}$ admit asymptotic power series as well, we write $|q_{N+1}(t;\epsilon)| = \epsilon^{N+1} C_{N+1}(\zeta, \epsilon)$. One verifies that $C_{N+1}(t,\epsilon)$ is bounded on any domain of the sort $(\zeta, \epsilon) \in (-\infty,\Theta] \times (0, \infty) \equiv \mathcal{D}(\Theta)$. This yields the uniform bound
\begin{equation}
\frac{1}{\epsilon^{N+1}} \left( \sum_{n > N} |p_{n}(\zeta;\epsilon)| + \sum_{n > N}|q_{n}(\zeta;\epsilon)| \right) \leq \exp\left(||C||_{L^{\infty}(\mathcal{D}(\zeta))} \int_{-\infty}^{\zeta} |\alpha(\zeta')| d \zeta'\right)< \infty. 
\end{equation}
Hence, $\sum_{n > N} p_{n}, \sum_{n > N}q_{n} \in O(\epsilon^{N+1})$. Summing the asymptotic expression for $p_{n}$ and $q_{n}$ therefore shows that $p$ and $\exp(\frac{2i}{\epsilon} \zeta)q$ admit asymptotic power series. Using the form of $\tilde{\psi}$ in the far past, we verify the real and imaginary parts of $\tilde{\psi}$ are linearly independent and normalized so that $\tilde{w}(t;\epsilon) = |\tilde{\psi}(t;\epsilon)|$ solves the deformed Ermakov-Pinney equation. Defining $w(t;\epsilon) = \tilde{w}(t;\epsilon)$ for $t \geq 0$ and uniquely extending $w(t;\epsilon)$ to $t < 0$, we have a solution to the Ermakov-Pinney equation \eqref{EPeqn} admitting an asymptotic power series of the form $w(t;\epsilon) \sim \Omega^{-1/2}(t) + \epsilon^2 w_{2}(t) + \hdots $ when $t \geq 0$. To show that $w(t;\epsilon)$ admits the same asymptotic series for $t < 0$, one can again use the Bremmer series with $\tilde{\Omega}(t)$ replaced by $\Omega(t)$ and with $p(0;\epsilon)$ and $q(0;\epsilon)$ giving the initial conditions at $t = 0$. The proof essentially proceeds as above. We remark that the cutoff function $H$ is only used to find suitable initial conditions for $w(t;\epsilon)$. 

To complete the proof that $w(t;\epsilon)$ is the desired function, we now show that $w'(t;\epsilon)$ also admits an asymptotic series. The proof is simple and leverages that $\tilde{w}(t;\epsilon)$ solves an initial value problem. Indeed, we have that
\begin{equation}
\frac{d}{dt}\tilde{w}(t;\epsilon) = \epsilon^{-2} \int_{-K}^{t} \left(\frac{1}{\tilde{w}^3(t',\epsilon)} - \tilde{\Omega}^2 \tilde{w}(t';\epsilon) \right)dt'.
\end{equation}
Using that $1/z^3$ is analytic away from $z = 0$, the integrand admits an asymptotic power series. Integrating term-by-term, we therefore have that
\begin{align}
\frac{d}{dt}\tilde{w}(t;\epsilon) &\sim \epsilon^{-2} \int_{-K}^{t} \left[\tilde{\Omega}^{3/2}(t')(- 3 \tilde{w}_{2}(t') \tilde{\Omega}^{1/2}(t')\epsilon^2 + \hdots) - \tilde{\Omega}^2(t') (\tilde{w}_{2}(t') \epsilon^2 + \hdots) \right] dt'
\\ &= a_{0}(t) + \epsilon^2 a_{2}(t) + \epsilon^4 a_{4}(t) + \hdots.
\end{align}
Since $\tilde{w}(t;\epsilon) = \Omega^{-1/2}_{0} + \int_{-K}^{t} \frac{d}{dt}\tilde{w}(t';\epsilon)dt'$, the uniqueness of asymptotic series implies that
\begin{equation}\label{integral}
\tilde{w}_{0}(t) + \epsilon^2 \tilde{w}_{2}(t) + \hdots = \Omega^{-1/2}_{0} + \int_{-K}^{t} \left( a_{0}(t') + \epsilon^2 a_{2}(t') + \hdots\right)dt'.  
\end{equation}
Since $\tilde{w}_{n}(t)$ are smooth, differentiating Eqn. \eqref{integral} gives for $t \geq 0$ that
\begin{equation}\label{asymototic}
\frac{d}{dt}w(t;\epsilon) \sim \frac{d}{dt} w_{0}(t) + \epsilon^2 \frac{d}{dt} w_{2}(t) + \hdots.
\end{equation}
Eqn. \eqref{asymototic} easily extends to $t < 0$ by only a trivial modification of the proof above. Hence, the asymptotic series for $w'(t;\epsilon)$ is obtained by term-by-term differentiation of the asymptotic series for $w(t;\epsilon)$. 

We note that if $\int_{-\infty}^{\infty} |\frac{d^{n}}{dt^{n}}\frac{\Omega'}{2 \Omega}| < \infty$ for all $n \geq 0$, then setting $H$ to unity shows $|w(\infty,\epsilon) - w(-\infty,\epsilon)| \sim 1 + 0 \epsilon + 0 \epsilon^2 + \hdots$. Hence $\mathcal{M}_{\epsilon}$ is conserved to all orders. This is essentially the result of Littlewood \cite{LITTLEWOOD1963233}. More refined error bounds for the adiabatic invariance of the simple harmonic oscillator are discussed in \cite{Invariance_of_SHO, doi:10.1137/0508051,doi:10.1137/S0036141097321516,NEISHTADT198158}. 

The Bremmer series is remarkable in that it \enquote{reorders} the formal sum of $\psi(t;\epsilon)$ to give a convergent solution $\psi = \psi_{0} + \psi_{1} + \hdots$ to Hill's equation with $|\psi_{n}| \in O(\epsilon^{n})$. Other methods of constructing convergent WKB-like solutions can be found in \cite{PhysRevD.72.104011,div_free_WKB,QLandWKB}. When $\Omega(t)$ is analytic, more refined methods based on Borel summation can be used to give exact solutions $\psi(t;\epsilon)$ with stronger asymptotic and analytic properties. We refer the interested reader to \cite{O_Costin_2004,asmtotitcs_and_borel_summability,ExactWKB,Voros,exact_WKB_2}.

\section{Section III}\label{sec3}
In this section, we fix $\epsilon$ and study the nonperturbative dynamics of the magnetic system when $\Omega^2(t)$ is $T$-periodic. Using Floquet theory, we identify a form for $w(t)$ which greatly simplifies and elucidates the long-time dynamics of particles. This section illuminates some possible pitfalls of using a perturbative theory when $\epsilon$ is not so small or when long time-scales are considered.

Consider Hill's equation in first-order form with 
\begin{align}\label{first_order_form}
   \frac{d}{dt} \begin{pmatrix} \psi \\ \varphi \end{pmatrix} = \epsilon^{-1} \begin{pmatrix} 0 & 1 \\ -\Omega^2(t) & 0 \end{pmatrix}\begin{pmatrix} \psi \\ \varphi \end{pmatrix} = \epsilon^{-1} M(t) \begin{pmatrix} \psi \\ \varphi \end{pmatrix}.
\end{align}
Let $\Phi(t;\epsilon^{-1}) = \mathcal{T} \exp(\epsilon^{-1} \int_{0}^{t} M(t')dt')$ be the fundamental matrix of Hill's equation, and $B(\epsilon^{-1}) = \Phi(T;\epsilon^{-1})$ the monodromy matrix. Let $\lambda_{1}, \lambda_{2}$ be the eigenvalues of $B(\epsilon^{-1})$, and suppose $(\mathbf{v}_{1}, \mathbf{v}_{2}) \in \mathbb{C}^2 \times \mathbb{C}^2$ is a basis putting $B$ in Jordan normal form. We define $\mathbf{z}_{1}(t) = \Phi(t;\epsilon^{-1})\mathbf{v}_{1}$ and $\mathbf{z}_{2}(t) = \Phi(t;\epsilon^{-1})  \mathbf{v}_{2}$. Trivially, $\mathbf{z}_{1}(t)$ and $\mathbf{z}_{2}(t)$ are linearly independent solutions to Hill's equations obeying $\mathbf{z}_{i}(t+T) = B \mathbf{z}_{i}(t)$. 

By Abel's theorem, $\det(B(\epsilon^{-1})) = 1$. If $\lambda_{1}= \lambda_{2}^*$ are non-real, then $\lambda_{i}$ are distinct and have unit modulus. It follows that all solutions to Hill's equation are bounded. Further, there exists a constant $c> 0$ such that
\begin{equation}
    w_{\text{per}}(t) = c |\mathbf{z}^{1}_{1}(t)|,
\end{equation}
is a periodic solution to the Ermakov-Pinney equation \footnote{One can verify this is the only $T-$periodic orbit.}. One easily computes that
\begin{equation}\label{scaling}
    c = \left|\mathbf{z}_{1}^{1}(0)\right|^{-1}\left|\text{Im}\left(\frac{\mathbf{z}_{1}^{2}(0)}{\mathbf{z}_{1}^{1}(0)}\right)\right|^{-1/2}.
\end{equation}
In the case when a periodic solution of Ermakov-Pinney exists, the exact invariant $\mathcal{M}$\footnote{We drop the subscript $\epsilon$ since $w_{\text{per}}(t)$ may not agree with $w(t;\epsilon)$ constructed in the previous section.} can be constructed to be time-periodic. Partially compactifying the particle phase space by quotienting $t \sim t + T$, one can verify the invariance of $(\mathcal{M}, \tilde{H}_{\epsilon}, p_{\theta})$ implies the existence of action-angle variables by the Arnold-Liouville theorem. This opens the door to a renewed study of more complex magnetic geometries through, for example, Hamiltonian perturbation theory \cite{Ham_perturbation_theory_1,Ham_pert_theory_2,Arnold_1989}.

When the $\lambda_{i}$ are strictly real, there is a constant $a>0$ such that 
\begin{equation}\label{general}
    w_{\text{can}}(t) = a |\mathbf{z}_{1}(t) + i \mathbf{z}_{2}(t)|
\end{equation}
solves the Ermakov-Pinney equation. If $\lambda_{1} = \lambda_{2} = \pm 1$ and $B$ is diagonalizable, then $w_{\text{can}}(t)$ is periodic and all solutions to Hill's equation are bounded. Otherwise, $w(t)$ grows unbounded for any choice of solution to the Ermakov-Pinney equation. It follows all particle trajectories grow unbounded at a typically exponential rate. This divergence may be attributed to the particle cyclotron motion resonating with the induced electric field. In such cases, it is not possible to define action-angle variables on the partially compactified particle phase space. 

Letting $\Delta(\epsilon^{-1}) = \text{tr}(B(\epsilon^{-1}))$, we have proved the long-term confinement of particles requires $|\Delta(\epsilon^{-1})| \leq 2$, with stability only ensured when the strict inequality holds. This fact could not have been uncovered at any order in perturbation theory. Perturbative theory predicts that $\mu_{\epsilon}$ is time-periodic to all orders. If the series for $\mu_{\epsilon}$ converged, the limiting invariant would be time-periodic as well. This would imply all particle trajectories are bounded. However, even small deviations of $\Omega(t)$ from a constant function may cause particles to become de-confined after some time. Indeed, suppose that $\Omega^2(t) = \Omega_{0}^2 - \epsilon^2 \phi(t)$. Then Hill's equation becomes the one-dimensional Schr\"odinger equation
\begin{equation}\label{SE}
    -\psi'' + \phi(t) \psi = E(\epsilon^{-1}) \psi, 
\end{equation}
with $E(\epsilon^{-1}) = \Omega_{0}^2\epsilon^{-2}$. For any fixed $\phi(t)$, the oscillation theory of Haupt \cite{hill1} shows there exists two strictly increasing sequences $\rho_{0} \leq \rho_{2} \leq \hdots \to \infty$ and $\rho'_{1} \leq  \rho'_{2} \leq \hdots \to \infty$ obeying the inequalities $\rho_{0} < \rho_{1}' \leq \rho_{2}' < \rho_{1} \leq \rho_{2} < \rho_{3}'\leq \rho_{4}' < \rho_{2} \leq \hdots$ such that the eigenvalues of $B(\epsilon^{-1})$ are non-real iff 
\begin{equation}\label{instab}
    E(\epsilon^{-1}) \in (\rho_{0},\rho_{1}') \cup (\rho_{2}', \rho_{1}) \cup (\rho_{2}, \rho_{3}') \cup \hdots.  
\end{equation}
Eqn. \eqref{instab} shows that, in general, infinitely many regions of instability are incurred as $\epsilon^{-1} \to \infty$. Regarding $\epsilon^2 \phi(t)$ as a perturbation, one should expect regions of instability to form around $\epsilon^{-1}$ values resonating with the unperturbed gyrofrequency, such as in Mathieu's equation \cite{Kovacic2018MathieusEA,Arn83,analytical_exp}.

 The stability properties of Hill's equation in the form of Eqn. \eqref{SE} is a classical and well-studied subject \cite{hill1,hill2, instab_intervals_of_Hills_Equation,geometric_aspects_of_stab,doi:10.1137/0145011}. Work on the asymptotic stability of Hill's equation in the more relevant form \eqref{first_order_form} has also been conducted by Weinstein and Keller in \cite{asym_hills_eqn}. Weinstein and Keller have shown that for a certain class of analytic $\Omega(t)$, the regions where $|\Delta(\epsilon^{-1})| \geq 2$ grow exponentially small as $\epsilon^{-1} \to \infty$. Hence, particle orbits are bounded for most $\epsilon \ll 1$. 
 
 We remark that when $|\Delta(\epsilon^{-1})| < 2$, the existence of an exact, time-periodic invariant proves the global error in the magnetic moment is uniformly bounded. This condition can be used to correct some of the statements in Qin and Davidson \cite{qin_exact_2006} when $\Omega(t)$ is periodic. 
\section{Section IV}\label{sec4}
We now use the results of the previous sections to compute and explore the accuracy of the adiabatic invariant $\mu_{\epsilon}$ when $\epsilon$ is not so small. By direct numerical computation, and by comparison to $\mathcal{M}_{\epsilon}$, we find that $\mu_{\epsilon}$ may not always be useful for studying the particle dynamics. We demonstrate how, with minimal computation, the methods developed in this work present a better alternative to the perturbative theory of confinement in time-varying magnetic fields.

For our example, we choose the $T = 2\pi$ periodic magnetic field
\begin{equation}
\Omega^{-1/2}(t) = (1 + \exp(\sin(t)).
\end{equation}
As we have fully justified, the formal solution to the Ermakov-Pinney equation,
\begin{equation}
w(t;\epsilon) = w_{0}(t) + \epsilon^2 w_{2}(t) + O(\epsilon^4) = \Omega^{-1/2}(t) - \frac{\epsilon^2 }{4 \Omega^2}\frac{d^2}{dt^2}(\Omega^{-1/2})(t) + O(\epsilon^4), 
\end{equation}
allows us to compute Kruskal's adiabatic invariant $\epsilon^{-1} \mu_{\epsilon} = \mu_{1} + \epsilon \mu_{2} + \epsilon^2 \mu_{3} + \epsilon^3 \mu_{4} + O(\epsilon^4)$. Namely, we have that 
\begin{align}
\epsilon^{-1} \mu_{1} &= \frac{ p_{\theta}^2}{4} \left(\frac{w^2_{0}(t)}{r^2} \right) + \frac{1}{4} \left(\frac{r^2}{w_{0}^2(t)} \right) + \frac{1}{4}(p_{r}w_{0}(t))^2 - \frac{p_{\theta}\sigma}{2}, \\
\epsilon^{-1} \mu_{2} &= -\frac{p_{r}r }{4} \frac{d w_{0}^2}{dt}(t),\nonumber \\
\epsilon^{-1} \mu_{3} &= \frac{p_{\theta}^2}{2} \left( \frac{w_{0}(t)w_{2}(t)}{r^2}\right) - \frac{r^2 w_{2}(t)}{2 w_{0}(t)^3}  + \frac{r^2}{4}\left( \frac{d w_{0}}{dt}(t)\right)^2 + \frac{p_{r}^2}{2} w_{0}(t) w_{2}(t), \nonumber \\ 
\epsilon^{-1} \mu_{4} &= - \frac{p_{r}r}{2} \frac{d}{dt}(w_{0} w_{2})(t). \nonumber
\end{align}

\begin{figure}[h!]
\centering
\begin{subfigure}{.5 \textwidth}
  \centering
  \includegraphics[width=\linewidth]{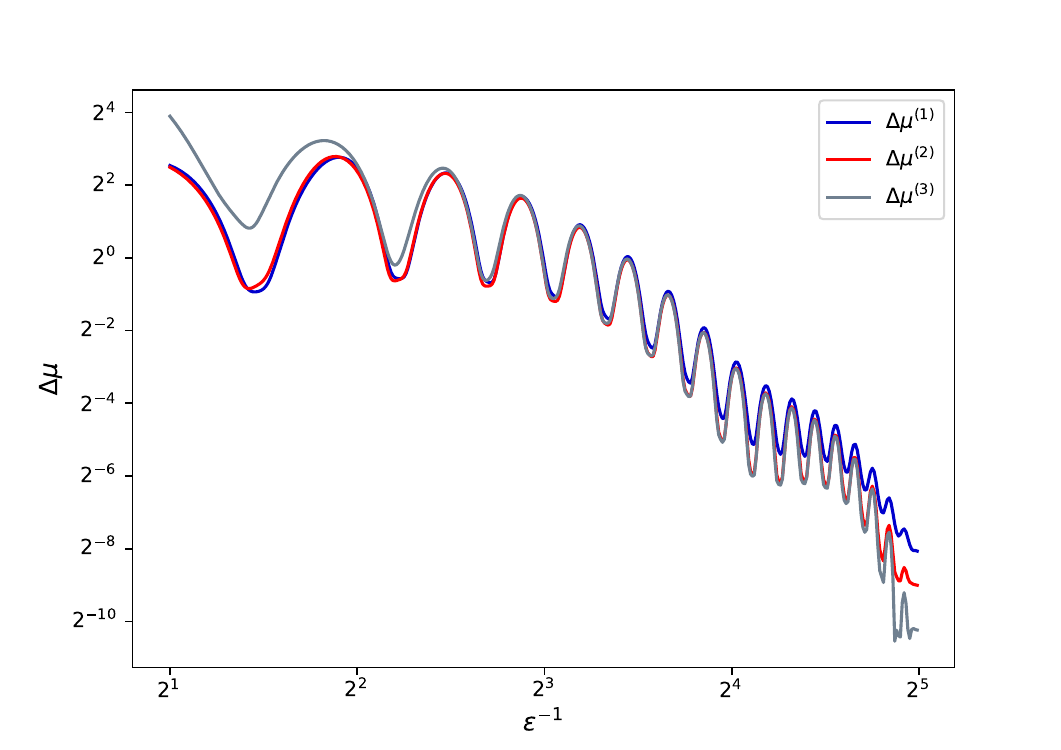}
\end{subfigure}%
\begin{subfigure}{.5 \textwidth}
  \centering
  \includegraphics[width=\linewidth]{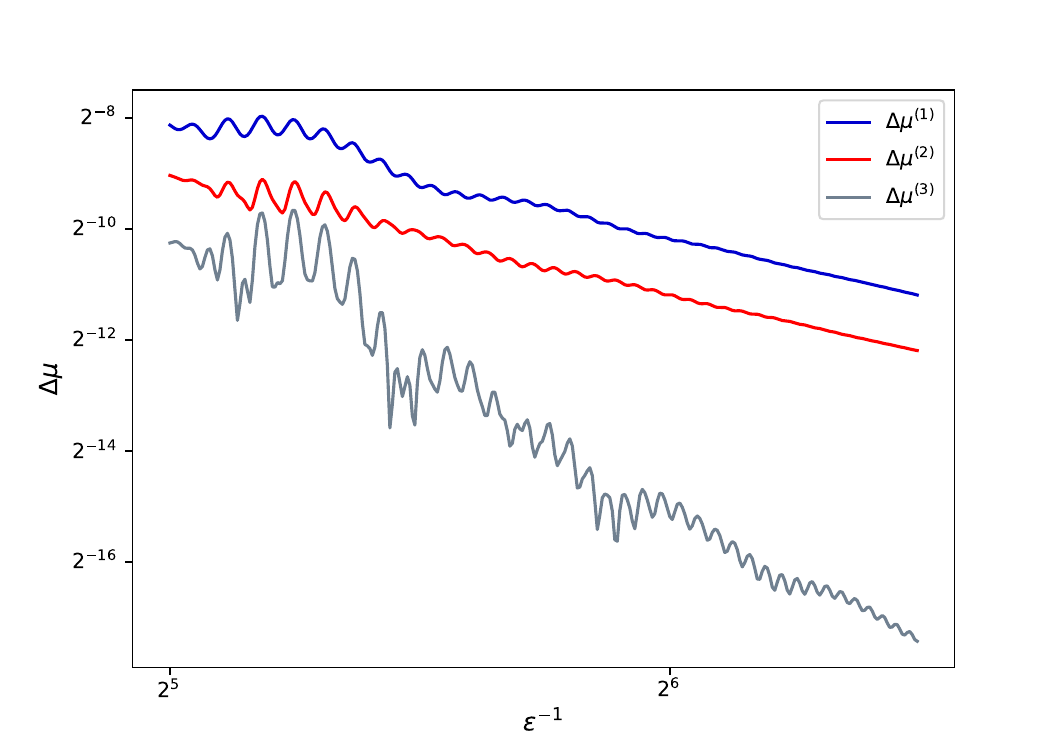}
\end{subfigure}
\caption{$\Delta \mu^{(n)}$ vs. $\epsilon^{-1}$ for the initial conditions $(r(0),p_{r}(0)) = (2,0)$. It is expected that $\Delta\mu^{(1)}_{\epsilon}$ and $\Delta\mu^ {(2)}_{\epsilon}$ are both $O(\epsilon^2)$ since the particle initially has no velocity.}
\label{fig1.0}
\end{figure}

\begin{figure}[h!]
\centering
\begin{subfigure}{.5 \textwidth}
  \centering
  \includegraphics[width=\linewidth]{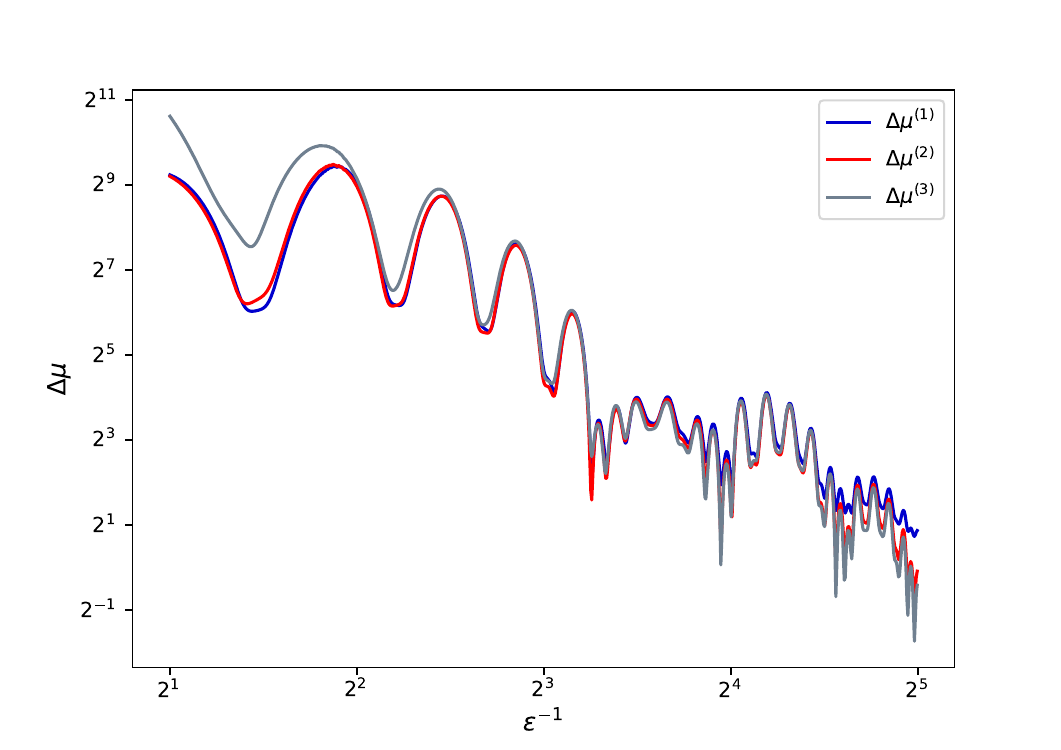}
\end{subfigure}%
\begin{subfigure}{.5 \textwidth}
  \centering
  \includegraphics[width=\linewidth]{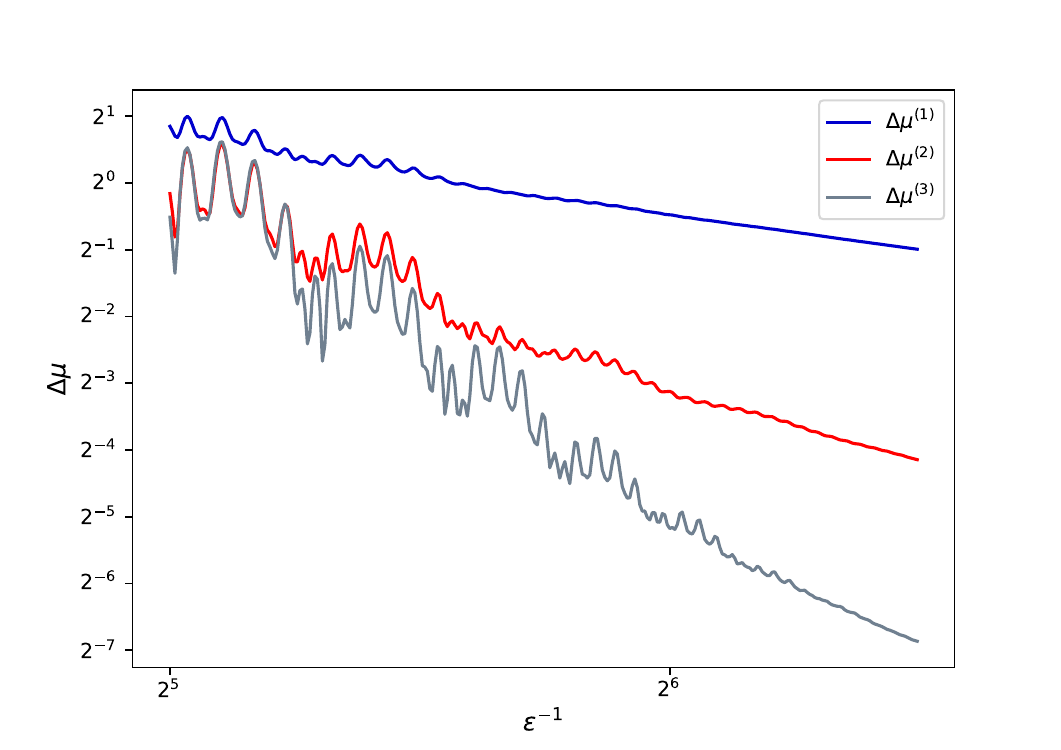}
\end{subfigure}
\caption{$\Delta \mu^{(n)}_{\epsilon}$ vs. $\epsilon^{-1}$ for the initial conditions $(r(0),p_{r}(0)) = (2,5)$. The typical asymptotic behavior as $\epsilon \to 0$ is seen for $\Delta \mu^{(n)}_{\epsilon}$.}
\label{fig1.1}
\end{figure}

\begin{figure}[h!]
\centering
\includegraphics[width=.7\linewidth]{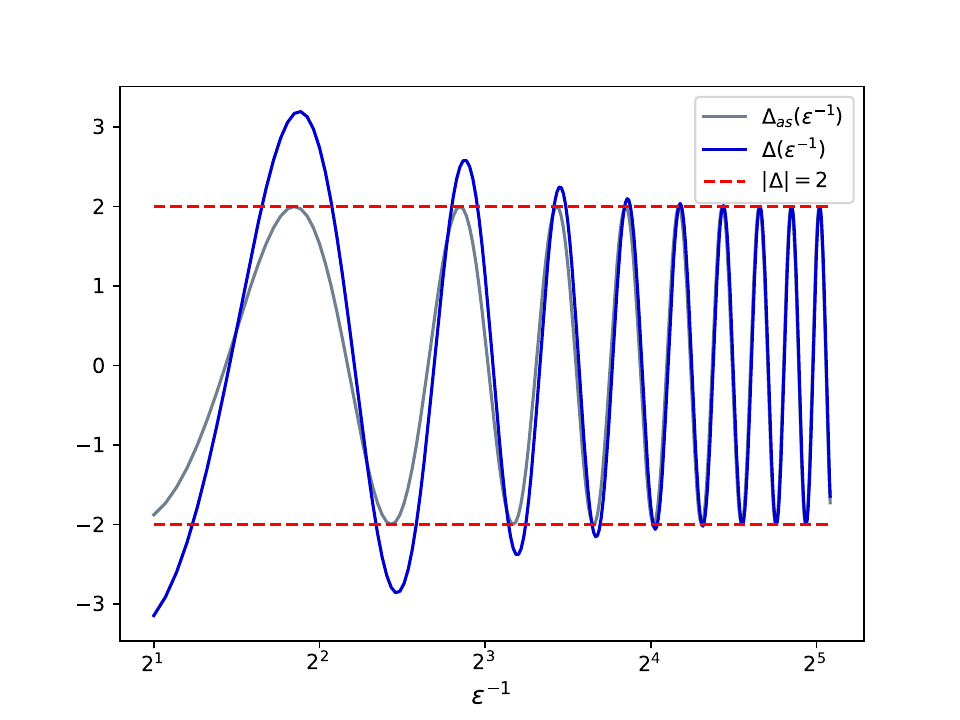}
\caption{$\Delta(\epsilon^{-1})$ obtained numerically and $\Delta_{\text{as}}(\epsilon^{-1})$ against $\epsilon^{-1}$ for relatively large $\epsilon$. The particle dynamics is stable when $|\Delta(\epsilon^{-1})|<2$.}
\label{fig2}
\end{figure}

\begin{figure}[h!]
\centering
\begin{subfigure}{.5 \textwidth}
  \centering
  \includegraphics[width=\linewidth]{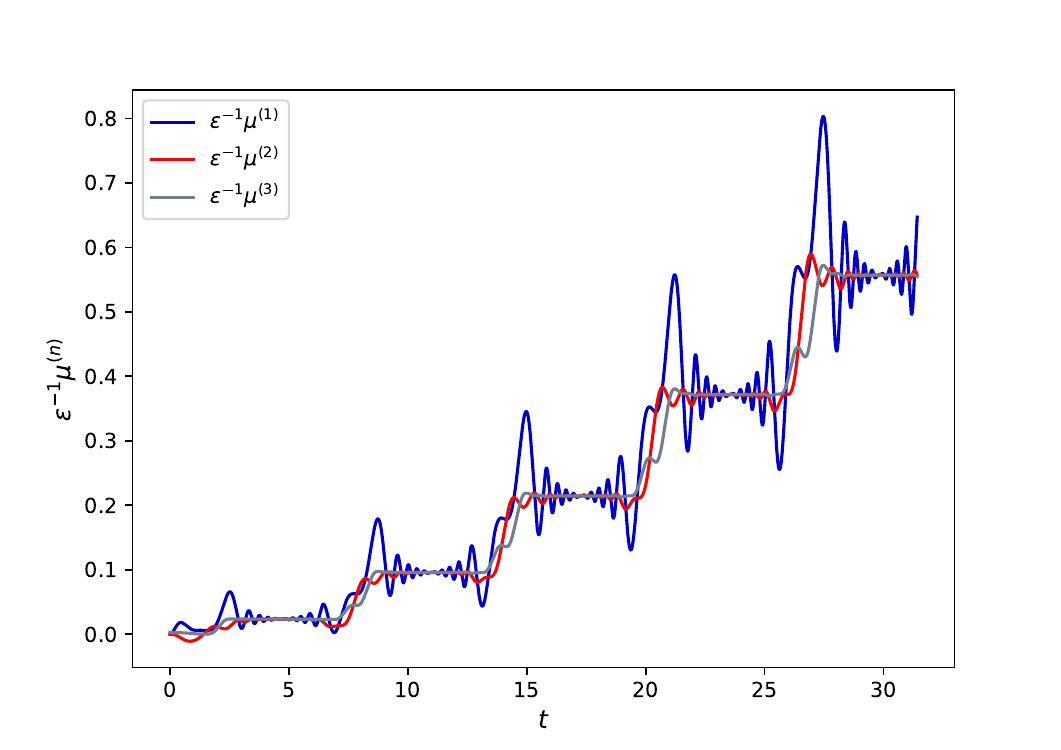}
\end{subfigure}%
\begin{subfigure}{.5 \textwidth}
  \centering
  \includegraphics[width=\linewidth]{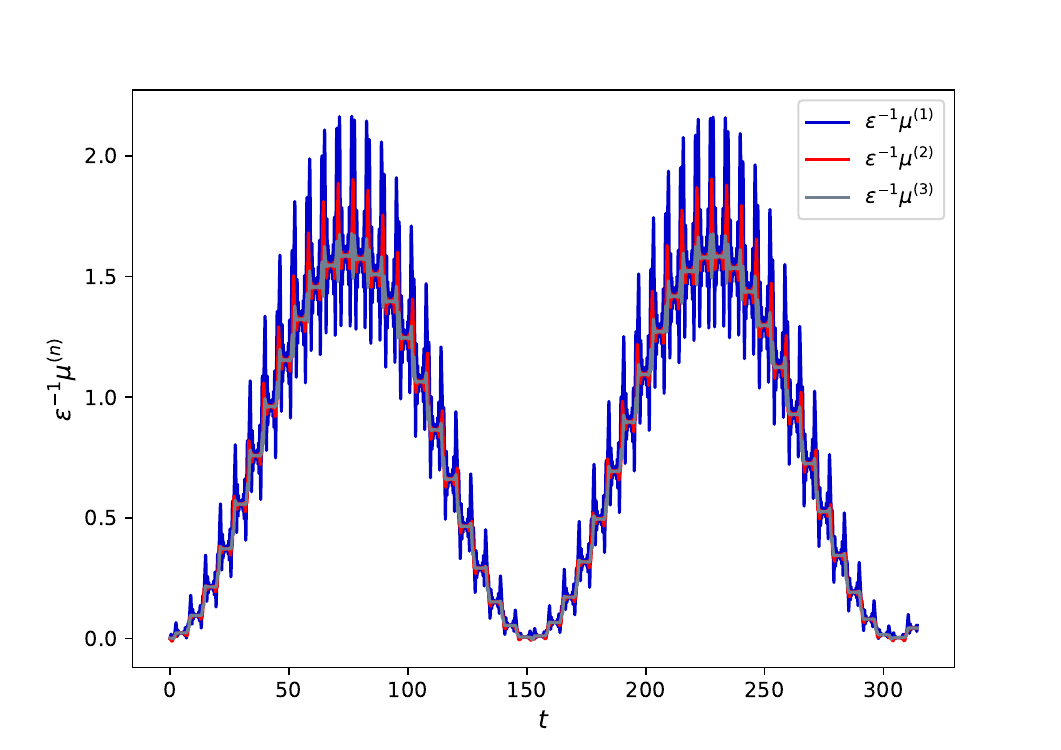}
\end{subfigure}
\caption{$\epsilon^{-1} \mu_{\epsilon}^{(n)}(t)$ vs. $t$ for the initial condition $(r(0),p_{r}(0)) = (2,0)$ and $\epsilon^{-1} = 20$.}
\label{fig3}
\end{figure}

To numerically explore the error of the conservation of the $n$-th order invariant $\epsilon^{-1} \mu^{(n)} = \mu_{1} + \hdots + \epsilon^{n-1}\mu_{n}$, we set $p_{\theta} = \sigma = 1$. We define the average $L^{2}$ error in the conservation of $\epsilon^{-1}\mu^{(n)}$ over two periods $\Delta \mu_{\epsilon} \equiv \frac{\epsilon^{-1}}{4 \pi}||\mu^{(n)}_{\epsilon}(r(t),p_{r}(t)) - \mu^{(n)}_{\epsilon}(r(0),p_{r}(0))||_{L^{2}(0,4\pi)}$. Using the initial conditions $(r(0),p_{r}(0)) = (2,0)$, corresponding to initially unmoving particles, and the initial conditions $(r,p_{r}) = (2,5)$, corresponding to initially gyrating particles, we simulate the particle orbits $(r(t),p_{r}(t))$  using Eqn. \eqref{EOM} and the optimal fourth order symplectic integrator of $\cite{symp_integrator}$. We then compute $\Delta \mu^{(n)}_{\epsilon}$ and plot the result against $\epsilon^{-1}$. The results are shown in Fig. \ref{fig1.0} and Fig. \ref{fig1.1}.
\FloatBarrier
\newpage
It is interesting that $\Delta \mu_{\epsilon}$ oscillates as it tends to zero with $\epsilon$. For any fixed, narrow range of $\epsilon$ values, the relative amplitude of these oscillations increases sharply as the accuracy of $\epsilon^{-1}\mu_{\epsilon}^{(n)}$ is computed over longer time intervals \footnote{The absolute size of these oscillations are nevertheless very small for small $\epsilon$.}. This is not unexpected. When $|\Delta(\epsilon^{-1})| > 2$, the resonance between the electric field and the cyclotron motion causes particle orbits to grow unbounded exponentially fast. This implies $\displaystyle\lim_{t \to \infty}|\mu_{\epsilon}(t) - \mu_{\epsilon}(0)| = \infty$ \footnote{This is not to imply that $\mu_{\epsilon}^{(n)}$ is not well conserved over long time intervals.}. For these resonant $\epsilon$ values, one therefore expects that $\epsilon^{-1}\mu_{\epsilon}^{(n)}$ is poorly conserved relative to the smallness of $\epsilon$. The locations of these ill-behaved $\epsilon$ values can be estimated using the zeroth-order WKB approximation. As a zeroth-order approximation, one has that
\begin{equation}
    \Delta(\epsilon^{-1}) = 2 A(\epsilon^{-1}) \cos \left(\epsilon^{-1} \int_{0}^{T} \Omega(t')dt' \right) \equiv A(\epsilon^{-1})\Delta_{\text{as}}(\epsilon^{-1}),
\end{equation}
for some function $A(\epsilon^{-1})$ limiting to unity as $\epsilon^{-1} \to \infty$. Arbitrarily long-term particle confinement may therefore fail when $\epsilon^{-1}\int_{0}^{T} \Omega(t')dt' \approx \pi N$ for some $N \in \mathbb{Z}$. For the present example, we compute 
\begin{equation}
    \int_0^{2\pi}\Omega(t^\prime)dt^\prime =  \int_{0}^{2 \pi} ( 1 + \exp(\sin(t'))^{-2}dt' \approx 1.7452.
\end{equation}
We numerically compute $\Delta(\epsilon^{-1})$ by finding the fundamental solutions to Hill's equation and compare the result to $\Delta_{\text{as}}(\epsilon^{-1})$.  The results are shown in Fig. \ref{fig2}. The oscillations in this plot can be directly compared to the oscillations in Fig. $\ref{fig1.0}$. 
\FloatBarrier
When $\epsilon^{-1}$ is not much bigger than $20$, Fig. \ref{fig1.0} and Fig. \ref{fig1.1} show that $\mu^{(3)}$ does not improve $\mu^{(2)}$ by much. Up to $O(\epsilon^2)$ terms, the approximate conservation of $\mu^{(2)}$ amounts to the conservation of
\begin{equation}
\epsilon^{-1} \tilde{\mu}^{(2)} = \frac{ p_{\theta}^2}{4} \left(\frac{\Omega^{-1/2}(t)}{r} \right)^2 + \frac{1}{4} \left(\frac{r}{\Omega^{-1/2}(t)} \right)^2 + \frac{1}{4}\left(p_{r}\Omega^{-1/2}(t) - \epsilon r\frac{d}{dt} \Omega^{-1/2}(t)\right)^2.
\end{equation}
For the present example, this should imply the $2\pi$-flowmap $\exp(2 \pi X_{\epsilon})$ has the approximate invariant 
\begin{equation}\label{pertubvative}
\epsilon^{-1} \tilde{\mu}_{\text{dis}}^{(2)}(\epsilon) = \frac{1}{r^2} + \frac{r^2}{16}  + \frac{1}{4}\left(2 p_{r}- \epsilon r \right)^2. 
\end{equation}
In general, nonperturbative theory implies that when $|\Delta(\epsilon^{-1})| < 2$, there exists an exact invariant of the $2\pi$-flowmap of the form 
\begin{equation}\label{discrete}
\epsilon^{-1} \mathcal{M}_{\text{dis}}(\epsilon) = \frac{1}{4} \left(\frac{a^2}{r^2} \right) + \frac{1}{4} \left(\frac{r^2}{a^2} \right) + \frac{1}{4}(p_{r}a - \epsilon b r)^2.
\end{equation}
Here, $a = w_{\text{per}}(0)$ and $b = w'_{\text{per}}(0)$ are the initial conditions of the periodic solution to the Ermakov-Pinney equation. In the present example, perturbative theory gives the approximation $a \approx 2$ and $ b \approx 1$. To obtain $a$ and $b$ exactly, one computes an eigenvector of the monodromy matrix $B(\epsilon^{-1})$ to Eqn. \eqref{first_order_form}. For example, if $\epsilon^{-1} = 20$ then 
\begin{equation}
B(20) =\begin{bmatrix}-0.8221 & -0.9786 \\ 0.0464 & -1.1613 \end{bmatrix}.
\end{equation}
$B(20)$ has strictly complex eigenvalues and has the eigenvector $
\mathbf{v} = \left( 0.9771, 0.1694 + 0.1286 i \right)^{t}$. Using Eqn. \eqref{scaling} to find the correct scaling for $\mathbf{v}$, we have exactly that $a = 2.756$ and $b = 20 \cdot 0.4777$. This differs considerably from the perturbative prediction. Evidently, $\mu_{\epsilon}$ fails to be well conserved. This becomes more evident when one looks at $\mu^{(n)}_{\epsilon}(t)$ in Fig \ref{fig3}. The large growth of $\mu_{\epsilon}^{(n)}$ reflects that the eigenvalues of $B(20)$ lie near the real axis, implying $\epsilon^{-1} = 20$ is nearly resonant. Since $\epsilon^{-1} = 20$ is not resonant, the Poincar\'e return theorem ensures that $\mu^{(n)}_{\epsilon}(t)$ returns close to its initial value. 

\begin{figure}[h!]
    \centering
    \includegraphics[width=.8\linewidth]{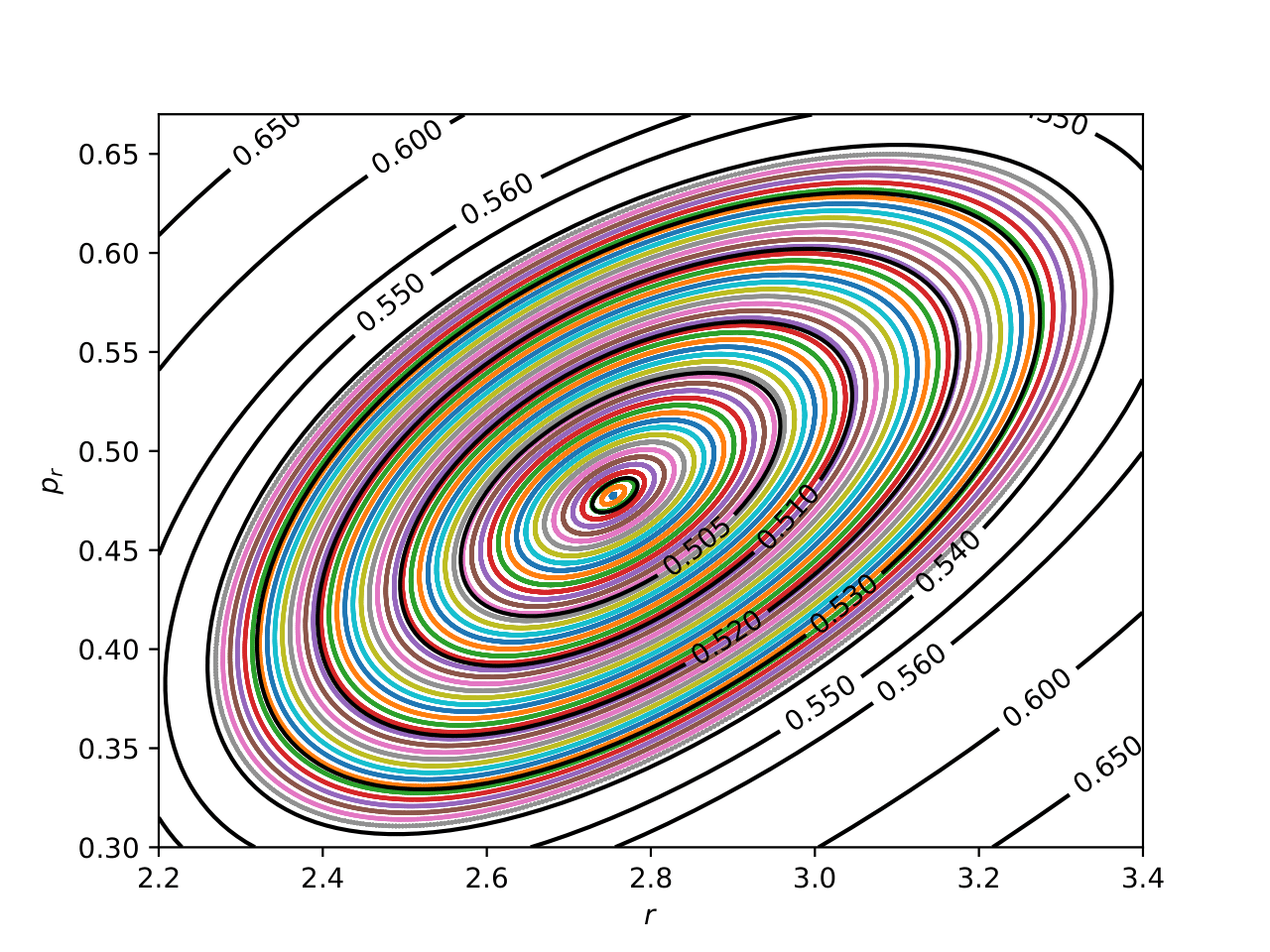}
    \caption{Poincaré plot of $2\pi$-flowmap near the fixed point. Level sets of $\mathcal{M}_{\text{dis}}(\epsilon)$ are overlayed in black. Particle orbits are obtained by a full-orbit simulation over 3000 periods.}
    \label{fig4}
\end{figure}

In contrast to the perturbative invariant \eqref{pertubvative}, the nonperturbative invariant $\eqref{discrete}$ fully characterizes the orbit structure of the $2\pi$-flowmap. We demonstrate this by running a full-orbit simulation of $(r(t),p_{r}(t))$ and plotting the resulting Poincaré plot. The result is shown in Fig. \ref{fig4} with the level sets of Eqn. \eqref{discrete} overlayed. We note that a full-orbit simulation is unnecessary beyond an independent validation of our results since $(r(2 \pi N), p(2 \pi N))_{N \in \mathbb{Z}}$ can be readily computed from $B(20)$. 
\FloatBarrier
\section{Discussion}\label{discussion}
In this work, we have carefully argued that two, seemingly contradictory, conclusions can be drawn about particle motion in a homogeneous, time-dependent magnetic field. The first is that there exists an exact invariant of motion $\mathcal{M}_{\epsilon}$ asymptotic to Kruskal's adiabatic invariant $\mu_{\epsilon}$ to all orders in $\epsilon$. The second is that, despite this fact, the exact charged particle dynamics may differ qualitatively from the approximate charged particle dynamics. These two facts coexist because, in general, adiabatic invariants are only approximately conserved over some finite (but long) time interval. This is evident from the system considered in this work, where $\epsilon$ values for which particle trajectories grow unbounded may accumulate around $\epsilon = 0$. We therefore emphasize that appropriate care must be taken when discussing perturbative and nonperturbative invariants, especially when $\epsilon$ is not so small, or when particle orbits grow unbounded. Such subtleties have not presented in previous work into the nonperturbative guiding center model, since only magnetic systems for which particle trajectories lie on invariant tori have been considered. We therefore hope this work serves as a roadmap to the analysis of more general nonperturbative models.

Many questions remain about $\mathcal{M}_{\epsilon}$, the solutions the the singularly-perturbed Ermakov-Pinney equation, and non-perturbative adiabatic invariants in general. One question of interest is the global error in $\mu_{\epsilon}^{(n)}$ when particle motion is uniformly bounded, and how this error grows as $\epsilon$ approaches an unstable value. Related to this question is exactly how long the usual magnetic moment can be accurate when particle trajectories are unbounded. 
Another question is how well the solutions to the Ermakov-Pinney equation constructed using the Bremmer series approximate the periodic solution for stable $\epsilon$ values and periodic $B(t)$. We have shown that $\mathcal{M}_{\epsilon}$ is asymptotic to the time-periodic adiabatic invariant series $\mu_{\epsilon}$ when the Bremmer-series solution is used to obtain $w(t;\epsilon)$, but not when the periodic solution is used. Another interesting study would be to compare the particle dynamics between stable and unstable $\epsilon$ values, and the way particle trajectories transition from bounded to unbounded. Of particular interest are the particle dynamics in a magnetic field of the form
\begin{equation}
    B(t) = \bar{B}(t) + \tilde{B}(t) 
\end{equation}
with $\tilde{B}$ a small, rapidly fluctuating magnetic field and $\bar{B}(t)$ a slowly-varying background magnetic field. 

We comment that it is likely impossible to define an exact nonperturbative invariant for most magnetic geometries. The magnetic fields considered in this work and in Hollas et al. \cite{Hollas} possess sufficiently many symmetries to ensure the integrability of the particle dynamics. It is currently unclear which magnetic geometries possess exact adiabatic invariants, but it is unlikely that exact nonperturbative invariants exist in most non-symmetric fields. We nevertheless conjecture that a satisfactory theory of approximate and/or non-smooth invariants may be possible. Irrespective of whether an analytical theory of approximate nonperturbative invariants can be developed, data-driven nonperturbative invariants already offer a powerful extension of guiding-center theory. Formulas for exact nonperturbative adiabatic invariants provide a unique opportunity to benchmark the learning error in such models. In a future work, we will use the methods of Burby et al. \cite{j_w_burby_nonperturbative_2025} to reconstruct the particle dynamics considered in this work. We will then compare the learned $\mathcal{M}_{\epsilon}$ against the exact $\mathcal{M}_{\epsilon}$. 

\section{Acknowledgments}
\noindent The first author was supported under DOE Contract No. DE-AC02-09CH11466. 
\\\\
This material is based on work supported by the U.S. Department of Energy, Office of Science, Office of Advanced Scientific Computing Research, as a part of the Mathematical Multifaceted Integrated Capability Centers program, under Award Number DE-SC0023164. It was also supported by U.S. Department of Energy grant \# DE-FG02-04ER54742.

\section{References}
\bibstyle{unsrt}
\bibliography{bib_file.bib}

\end{document}